\makeatletter
\newif\if@restonecol
\makeatother

\documentclass[runningheads]{llncs}

\input{psfig.sty}

\usepackage{times}
\usepackage{mathrsfs}
\usepackage{amssymb}
\usepackage[ruled]{algorithm2e}
\usepackage{graphicx}
\usepackage{savesym}
\usepackage{amsmath}
\savesymbol{iint}
\usepackage{txfonts}
\restoresymbol{TXF}{iint}
\begin{document}
\title{Incremental Collaborative Filtering Considering Temporal Effects}

\author{Yongji Wang\inst{1,2}
\and Xiaofeng Liao\inst{4,1,3} \and Hu Wu\inst{1} and Jingzheng Wu\inst{1,3} }

\authorrunning{Yongji Wang et al.}

\institute{National Engineering Research Center for Fundamental
Software, Institute of Software,
\\Chinese Academy of Sciences, Beijing, China\\
State Key Laboratory of Computer Science, Institute of Software,
\\Chinese Academy of Sciences, Beijing, China\\
Graduate University,Chinese Academy of Sciences, Beijing, China\\
Information Engineering School, Nanchang University, Nanchang, Jiangxi, China
\email{\{xiaofeng08\}@iscas.ac.cn}\\
}

\maketitle
\thispagestyle{empty}
\begin{abstract}
Recommender systems require their recommendation algorithms to be
accurate, scalable and should handle very sparse training data which
keep changing over time. Inspired by ant colony optimization, we
propose a novel collaborative filtering scheme: Ant Collaborative
Filtering that enjoys those favorable characteristics above
mentioned. With the mechanism of pheromone transmission between
users and items, our method can pinpoint most relative users and
items even in face of the sparsity problem. By virtue of the
evaporation of existing pheromone, we capture the evolution of user
preference over time. Meanwhile, the computation complexity is
comparatively small and the incremental update can be done online.
We design three experiments on three typical recommender systems,
namely movie recommendation, book recommendation and music
recommendation, which cover both explicit and implicit rating data.
The results show that the proposed algorithm is well suited for
real-world recommendation scenarios which have a high throughput and
are time sensitive.
\end{abstract}

\section{Introduction}
Recommender systems help to overcome information overload by
providing personalized suggestions based
on the user's history/user's interest. Because recommender systems can increase user experience
by providing more relative information or even find information user
can't find otherwise, they are deployed on many websites, especially
e-commerce websites \cite{E-Commerce} . According to relying on the
content of the item to be recommended or not, the underlying
recommendation algorithms are generally classified into two main
categories: content based and collaborative filtering (CF)
recommendations. In this paper, we focus on the collaborative
filtering algorithms. In the essence, they make recommendations to
the current user by fusing the opinions of the users who have
similar choices or tastes. It is a more natural way of
personalization because people are social animals. There are always
persons who share common interests with us that reliable
recommendations can be made upon.

The problem settings of collaborative filtering are simply described
as follows. The user set is denoted as $U$ and the item set is
denoted as $I$. Users give ratings $r \in [1, Max]$ to the items
that they have seen indicating their preference. So the ratings form
a rating matrix $R^{|U|\times|I|}$ where the unknown ratings are
left as $0$. Each row of the rating matrix represents a user and
each column represents an item. The goal of collaborative filtering
is therefore to choose and recommend the items that the current user
would probably like most according to this rating matrix. The
existing CF algorithms are divided into two categories: memory-based
methods and model-based methods. Memory-based CF algorithms directly
use the rating matrix for recommendation. They predict the unknown
preference by first finding similar users or similar items denoted
as user neighbors and item neighbors, respectively; by fusing these
already known and similar ratings they can guess the unknown
ratings. On the other hand, model-based CF algorithms learn an
economical models representing the rating matrix. These models are
refereed to as user profile or item profile. Recommendation thus
becomes easy and intuitive on these lower dimension attributes.

Whilst considered to be one of most successful recommendation
methods, collaborative filtering suffers from two severe problems,
namely sparsity and scalability \cite{ActiveSpreading}. Please note
that for a single user it is impossible for her to rate all the
items and it is impossible for a single item been rated by all the
user either. Actually, most values in the rating matrix are unknown,
i.e., $0$. Since our recommendations are solely depending on this
very sparse rating matrix, how to leverage these data to generate
good recommendations is challenging. On the other hand, real-world
recommender systems often have millions of users and items. For many
recommendation algorithms, the training model needs hours even days
to be updated. Unchanging and outdated recommendations are likely to
disappoint our users. So we require the algorithms to be as fast as
possible in both training and recommendation phases.

In real world recommender systems, another practical but often
overlooked issue related to high quality recommendation is how to
consider the evolution of user interests over time. Take news
recommender systems such as Google personalized news
\footnote{http://news.google.com}
 for example, there are at least two reasons
to consider time effects in their recommendation algorithms
\cite{GoogleNews}. First, people always want to read the latest
news. To recommend a piece of news happened ten days ago is not
likely of equal interest as the news happened just now. (Using a
slicing time window to cut the old news off may be one trivial
solution, but obviously not the ideal solution.) More importantly,
people's tastes are always changing. A young man would like to see
recommendations about digital cameras if he plans to buy one. But
after he already owns it, he will not take interest in the
recommendations on buying a new digital camera. So time factors are of vital
importance for the success of recommender systems in many
applications, especially e-commerce, advertisement and news
services.

To recommend using these sparse and evolving preference data, we
propose a novel collaborative filtering algorithm named Ant
Collaborative Filtering (ACF) which is inspired by Ant Colony
Optimization algorithms. Similar to other swarm intelligence
algorithms, ACF could handle very sparse rating data by virtue of
pheromone transmission and is a natural extension of other CF
techniques in recommender systems in which preferences keep
changing. We make an analogy of users to ants which carry specific
pheromone initially. When the user rates a movie or simply reads a
piece of news on the web, our algorithm links the user and the item
by the mechanism of pheromone transmission from the user to the item
and vice versa. So the types of pheromone and their amounts
constitute a clear clue of historical preference and turn out to be
a strong evidence for finding similar users and items.

The remainder of this paper is structured as follows: we first
introduce some preliminaries. In Section \ref{section3} we present
our Ant Collaborative Filtering algorithm and in Section
\ref{section4} we improve this algorithm using dimension reduction
technique. In Section \ref{section5}, some related works are
described. We then report the experimental results on two different
datasets in Section \ref{section6}. Finally, we conclude the paper
with some future works.

\section{Preliminaries}
\label{section2}
\subsection{Rating-based vs. Ranking-based Recommendation}
The majority of collaborative filtering algorithms follows the rating
prediction manner, i.e., predicting the ratings for all the unseen
items for the current user and then recommend the items with the
highest prediction scores. Yet an alternative view of the
recommendation task is to generate a top $N$ list of items that the
user is most likely interested in, where $N$ is the length of final
recommendation list. In this regard, collaborative filtering can be
directly cast as a relevance ranking problem \cite{Ranking}. These
two types of algorithms are called rating-based and ranking-based
recommendations, respectively.

One of the disadvantages of rating-based CF algorithms is that they
can not make good recommendations in the situation of implicit user
preference data. We should notice that explicit user rating data
that rating-based algorithms rely on are not always available or are
far from enough. In most cases, users express their preference by
implicit activities (such as a single click, browsing time, etc.)
instead of giving a rating. As the consequence of lack of these
rating data, rating-based CF algorithms cannot work properly. In
this sense, ranking-based algorithms are as important as
rating-based algorithms, if not more important.

\cite{ItemTopN} proposed an item ranking algorithm by computing their
similarity with the items that the customers have already bought.
\cite{LearningRank} proposed a learning to rank algorithm that can
find a function $f: U \times V \rightarrow \mathbb{R}$ that
correctly ranks the items as many as possible for the users. For
every item pair $v_j$ and $v_l$, if user $u_i$ likes $v_j$ more than
$v_l$, we have $f(u_i, v_j) > f(u_i, v_l)$, otherwise the ranking
error increases. Although similar algorithms have been applied to
ranking problem for search engine, the size of both users and items
in recommender systems makes this personal ranking algorithm
intimidating for implementation.

In general, ranking-based CF hasn't been paid enough attention to in
academia although it is already popular in commercial recommender
systems compared to rating-based recommendation. We deem the reason
for this is that it is relatively difficult to define a proper loss
function as their rating prediction counterpart. In this paper, both
types of algorithms are concerned.

\subsection{Recommendation using Bipartite Graph}
\begin{figure}[h]
\centering
\includegraphics[width=85mm,height=60mm]{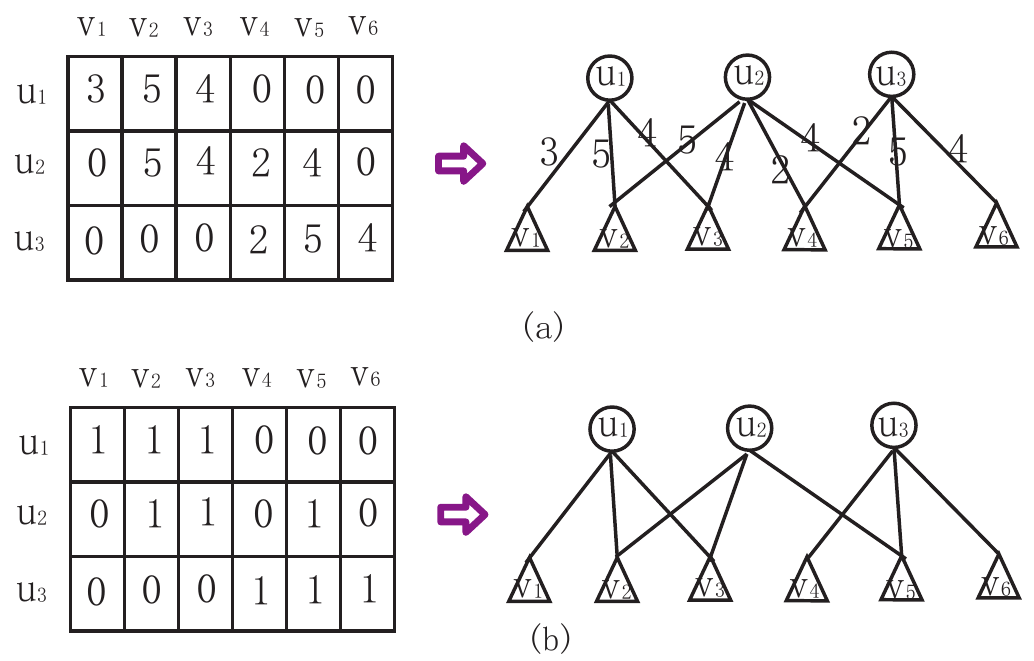}
\caption{(a) Explicit rating matrix and the corresponding weighted
bipartite graph, (b) Implicit preference matrix and the
corresponding unweighed bipartite graph} \label{fig:bipartite}
\end{figure}

Recommendation activity often involves two disjoint sets of
entities: the user set $U$ and the items set $V$. A natural
representation of these two groups is by means of bipartite graph
\cite{Bipartite}, where one set of nodes are items and the other set
are users. Links go only between nodes of different sets, in case
that the user selects the item or rates the item. The rating matrix
of collaborative filtering domain therefore could be elegantly
represented by this weighed undirected bipartite graph and the
original matrix is actually the adjacency matrix of this graph. As
for implicit preference matrix, the generated bipartite graph is
unweighted and undirected. The two scenarios are shown in Fig.
\ref{fig:bipartite} (a) and (b).

For rating prediction problem, we are interested in finding the user
neighbors and item neighbors in both modes of the graph
simultaneously. For ranking-based recommendation, we are interested
in finding the missing edges between the two partite indicating
potential interests between user and item. Hereafter, we cast the CF
problem as bipartite graph mining problem without special
explanation.

\subsection{Considering Time effects}
The data in recommender systems keep changing. Not only new users
and new items are continuously added to the system, the preferences
of existing users and the features of existing items are also
changing over time. To make better recommendations, algorithms
should update the learned model efficiently and appropriately. By
``efficiently'' we mean the algorithm has the ability for fast or
even real-time update. On the other hand, ``appropriately'' means we
must take time factor into consideration when deciding on what to
recommend to the users. For example, we can not expect users are
equally satisfied with the recommendations of a very old movie and a
brand new movie, even they have similar ratings.

In spite of its importance, time factors gain little attention in
recommendation research area until recently. The authors of
\cite{Proximity} proposed an incremental update method to compute
the proximity of any two given nodes in the bipartite graph changing
on the year base. \cite{Temporal} analyzed the evolution of ratings
in the Netflix movie recommender system. It proposed two CF
algorithms that consider time effects: item-item neighborhood method
and rating matrix factorization method. Both the methods are
elaboratively tailored for Netflix movie rating data and achieve the
best results so far.

As pointed out in \cite{Temporal}, preference evolution is subtle
and delicate. For a single user, what we can use are often just a
few preference instances inundated by millions of non-relevant
data. The above mentioned methods have been designed and tested for
special recommendation scenarios. However, in a more general sense,
a robust recommendation algorithm considering time evolution for a
wider range of applications including both rating-based and
ranking-based recommender systems is still absent and is the main
focus of our research.

\section{Ant Collaborative Filtering}
\label{section3} Before proceeding on a more detailed description,
we first introduce notations and  Ant Colony Algorithm which sheds
some light on our proposed methods.

\begin{table}[h]
\scriptsize \small \caption{Notations} \label{tab:notations}
\begin{center}
\begin{tabular}{p{1.5cm}p{6cm}} \hline
\textbf{Notation} & \textbf{Meaning} \\ \hline
 $u_i$, $v_j$ & A user and an item \\
 $r_{i,j}$ & The rating that user $u_i$ gives to item $v_j$, $0$ if unknown \\
 $\overline{r}$, $\overline{r}_{u_i}$, $\overline{r}_{v_j}$ & The average rating for all the ratings, user $u_i$
 and item $v_j$ \\
 $\mathscr{K}$, $K$ & Total number of types of pheromone and maximum
 of types of pheromones attached to a user or an item, $\mathscr{K} \geq
 K$ \\
 $C(u_i)$ & Neighborhood user set for user $u_i$\\
 $C(v_j)$ & Neighborhood item set for item $v_j$\\
 $\mathbf{Ph(u_i)}$, $\mathbf{Ph(v_j)}$ & The pheromones attached to user $u_i$ and item $v_j$, they are vectors of pheromones\\
 $\gamma$, $\lambda$, $\sigma$ & Parameters in our model,
 controlling the transmission rate, evaporating rate and
 disappearing rate \\
 $TopN$ & The recommenced top $N$ items \\
 $Hitting$ & The subset that user likes in the recommended list \\
 $rank(v_j)$ & The final rank position for item $v_j$ in rank-based recommendation \\
 \hline
\end{tabular}
\end{center}
\end{table}

\subsection{Ant Colony Algorithm}
Ant Colony Optimization (ACO) algorithms are proposed after
observing the automatic accumulation or communication phenomena
common to ant colony which thereafter have been widely applied in
various domains such as clustering and information retrieval
\cite{ACO}. They belong to a more general group of simulation
algorithms named swarm intelligence algorithms. In the typical
settings of ACO, every ant is identified by some kind of indicators
for communication often referred to as \textit{pheromone}. The
pheromone is used as an indirect communication medium. Taking
Traveling Salesman Problem (TSP) which finds the shortest path on a
graph for an example, while each ant walks on the graph, it leaves a
pheromone signal through the path it used. Shorter paths will leave
stronger signals. The next ants, when deciding which path to take,
tend to choose paths with stronger signals with a higher
probability, so that shorter paths are found.

ACO has several enticing properties. The most prominent one is that
they are dynamic and self-organizing in nature. So it suits to solve
the problems that are too complex and dynamic to be solved using
other machine learning methods directly optimize some utility
function.

There are applications of ACO in many information retrieval domains
such as web search and social network. For example, \cite{AntWeb}
proposed a framework that models the web surfing activity as an ant
colony group behavior. They took an analogy of users as ants and web
pages as food. Similar analogy can be mapped onto users and items in
the CF context.

\subsection{Ant Collaborative Filtering}
\begin{figure}[h]
\centering
\includegraphics[width=80mm,height=35mm]{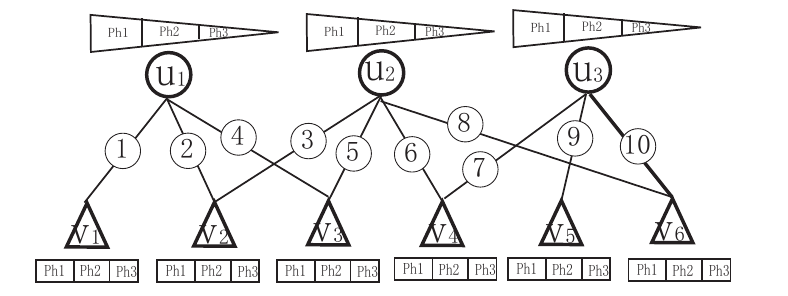}
\caption{Pheromone transmission between users and items after
several steps, the numbers in the circle indicate the timestamp, the
triangles contain user pheromones and the rectangles contain item
pheromones} \label{fig:pheromone}
\end{figure}

The intuition of our Ant Collaborative Filtering (ACF) algorithm is
that given pheromone representing a user or a group of users, the
item shares the user's pheromone when she rates the item. Meanwhile,
item transfers the pheromone already attached on it to the user. So
after some time, similar items receive similar pattern of pheromone
and then users with similar tastes become alike in respect to the
pattern of pheromone on them. The pheromone transimission process is
illustrated in Fig. \ref{fig:pheromone}, note that the ratings are
learnt one by one in their original time order. Thus far,
recommendation can be generated after two strategies: (1) Provided
similar users and similar items found, we can estimate the current
user's rating on the items that she hasn't seen before by simply
employing memory based CF methods; (2) We can rank the items
according to the similarity of its pheromones to the user's. These
two strategies correspond to rating-based and ranking-based
recommendation mentioned above, respectively.

\subsubsection{Training}
The training process is relatively simple and intuitive. Partial
reason is that we need to retrain the model whenever there are new
rating data available. In other words, our training process is
incremental and is completed online.

First we initialize user pheromone by allocating every single user a
unique pheromone with value $1.0$. The item pheromone is empty for
every item. When a user $u_i$ gives a rating $r_{i,j}$ to item
$v_j$, they exchange their pheromones, i.e., the user updates her
pheromone by adding the item's pheromones times by rating adjustment
and a constant $\gamma$ which is introduced to control the spreading
rate. For the item, similar update is computed. Because the bigger
the gap between the rating and the average rating is, the stronger
it bears user's preference. If the rating is much higher than the
average, the transmission will be a strong positive ``plus''; on the
other hand, if the rating is much less than the average, the
transmission will be a strong negative ``minus'' which means the
user and the item are not alike. So there are negative values for
the pheromones.

As mentioned before, user's interests change over time. Old
interests fade out and new interests develop. We capture this
evolution by the mechanism of pheromone evaporation. Before
pheromone exchange between the user and the item, existing
pheromones evaporate in a rate according the ratio of their amounts
to the highest concentration among all the existing pheromones.

Therefore, for the rating-based recommendation scenario, the
complete pheromone update formulae for the item and the user is
described in Equ. (1) and Equ. (2).

\begin{equation}
\label{eq:rating_pv}
\mathbf{Ph_{v_j}^{(t+1)}} = \mathbf{Ph_{v_j}^{(t)}} \times
\exp\left(\frac{amount_{v_j,k} + \lambda}{Max_{k \in
K}(amount_{v_j,k}) +
 \lambda} - 1\right)  \\ +
 (r_{i,j} - \overline{r}_{u_i}) \times \gamma \times
 \mathbf{Ph_{u_i}^{(t)}}
\end{equation}
\begin{equation}
 \mathbf{Ph_{u_i}^{(t+1)}} = \mathbf{Ph_{u_i}^{(t)}} \times \exp\left(\frac{amount_{u_i,k} + \lambda}{Max_{k \in K}(amount_{u_i,k}) + \lambda} - 1\right) \\ +
 (r_{i,j} - \overline{r}_{v_j}) \times \gamma \times \mathbf{Ph_{v_j}^{(t)}}
\end{equation}

For the 0/1 preference data, similarly, we rewrite the pheromone
update formula as follows:
\begin{equation}
\mathbf{Ph_{v_j}^{(t+1)}}=\mathbf{Ph_{v_j}^{(t)}} \times
\exp\left(\frac{amount_{v_j,k} + \lambda}{Max_{k \in
K}(amount_{v_j,k}) + \lambda} - 1\right) \\+ \gamma \times
\mathbf{Ph_{u_i}^{(t)}}
\end{equation}
\begin{equation}
\mathbf{Ph_{u_i}^{(t+1)}}=\mathbf{Ph_{u_i}^{(t)}} \times
\exp\left(\frac{amount_{v_j,k} + \lambda}{Max_{k \in
K}(amount_{u_i,k})+\lambda} - 1\right) \\ +  \gamma \times
\mathbf{Ph_{u_i}^{(t)}}
\end{equation}

After evaporation and transmission, we delete the pheromones whose
amount is less than our threshold $\sigma$ to keep our model simple
and robust to rating noise. This process is referred to as threshold
cut off. In our experiment, $\sigma$ is set to $0.01$. The
training algorithm is shown below.

\begin{algorithm}[!h]
\SetAlgoLined
\KwIn{$Ratings$, $U$, $V$} \KwOut{Updated user pheromone
and item pheromone}
 \caption{Training phase of ACF}
 \label{algo:aca_training}
 //Initialization \\
 \For {$u_i \in U$ } {
  $Ph_{u_i}^{(0)} = \{u_i:1.0\}$;
 }
 \For {$v_j \in V$} {
 $Ph_{v_j}^{(0)} = \{\}$;

 }
 //Training \\
 \For{$rating \in Ratings$} {
   $u_i = rating[user]$; \\
   $v_j = rating[item]$; \\
   $value = rating[value]$; \\
     // Update user pheromones \\
     \For {$Pheromone \in \mathbf{Ph_{u_i}^{(t)}}$} {
       // Evaporation and Transmission \\
       $Pheromone = Pheormone \times
       \exp(\frac{amount_{Pheromone} + \lambda}{Max_{Ph_{u_i}} + \lambda} - 1) +  \gamma \times (value - \overline{r}_{v_j}) \times Ph_{v_j}^{(t)}$
       \\ // Cut off \\
       \If{$abs(Pheromone) < \sigma$} {
         $Pheromone = 0$;
       }
       $\mathbf{Ph_{u_i}^{(t+1)}} \leftarrow Pheromone$;
     }
     // Update item pheromones \\
     \For {$Pheromone \in \mathbf{Ph_{v_j}^{(t)}}$} {
       // Evaporation and Transmission \\
       $Pheromone = Pheromone \times
       \exp(\frac{amount_{Pheromone} + \lambda}{Max_{Ph_{v_j}}+\lambda} - 1) + \gamma \times (value - \overline{r}_{u_i}) \times Ph_{u_i}^{(t)}$
        \\
       // Cut off \\
       \If{$abs(Pheromone) < \sigma$} {
         $Pheromone = 0$;
       }
       $\mathbf{Ph_{v_j}^{(t+1)}} \leftarrow Pheromone$;
     }

 }

\end{algorithm}

\subsubsection{Recommendation}
We give recommendation algorithms for both explicit rating
prediction and implicit relevance ranking tasks.

We can calculate the following three types of similarities though
pheromone comparison:
\begin{itemize}
\item \textbf{User-User similarity}, $\mathbf{s(u_i, u_k)}$: to what extent any two given users $u_i$ and $u_k$ are
alike, computed by comparison the user pheromones;
\item \textbf{Item-Item similarity}, $\mathbf{s(v_j, v_l)}$: to what extent any two given items $v_j$ and $v_l$ are alike, computed by comparison the item pheromones;
\item \textbf{User-Item similarity}, $\mathbf{s(u_i, v_j)}$: to what extent user $u_i$ and item $v_j$ are alike, computed by comparison the corresponding user and item
pheromones.
\end{itemize}

For rating prediction, we employ memory based methods, i.e., using
ratings from similar users to similar items to predict the current
user $u_i$'s rating on current item $v_j$. The central problem is to
find the best user neighbors and item neighbors that have similar
rating patterns. In our experiments, the neighbor number is fixed as
$20$ for both user neighbors and item neighbors. We first give
rating-based recommendation in Alg. \ref{algo:aca_rating}.

\begin{algorithm}[!h]
\SetAlgoLined
\KwIn{$u_i, v_j, Ph(U), Ph(V)$} \KwOut{$r_{i,j}$}
 \caption{Rating-based recommendation}
 \label{algo:aca_rating}
 // Find user neighbors \\

  \For {$user \in U$} {
    $s_{u_i, user} = \frac{\mathbf{Ph_{u_i}} \circ
    \mathbf{Ph_{user}}}{|\mathbf{Ph_{u_i}}| \times |\mathbf{Ph_{user}}|}$;

  }
  // Sort users according to their similarity to the current // user in descendant order\\
  $\mathbf{users} = sort(s_{u_i, user})$; \\
  $C(u_i) = \mathbf{users}[0:Neighborhood\_size]$; \\
 // Find item neighbors \\
 \For {$item \in I$} {
    $s_{v_j, item} = \frac{\mathbf{Ph_{v_j}} \circ
    \mathbf{Ph_{item}}}{|\mathbf{Ph_{v_j}}| \times |\mathbf{Ph_{item}}|}$;

  }
  // Sort items according to their similarity to the current // item in descendant order\\
  $\mathbf{items} = sort(s_{v_j, item})$; \\
  $C(v_j) = \mathbf{items}[0:Neighborhood\_size]$; \\
 // Fusing ratings \\
 $user\_prediction = 0;$ \\
 $user\_similarity = 0;$ \\
 \For {$user \in C(u_i)$} {
  \If {$user\ has\ rated\ item\ v_j$} {
    $user\_prediction\ +\hspace{-0.1cm}= abs(s(u_i, user)) * (r_{user, v_j} -
    \overline{r}_{user})$;
    $user\_similartity\ +\hspace{-0.1cm}= abs(s(u_i, user))$;
  }
  $user\_prediction\ /\hspace{-0.1cm}= user\_similarity$;
 }
 $item\_prediction = 0;$ \\
 $item\_similarity = 0;$ \\
 \For {$item \in C(v_j)$} {
  \If {$u_i\ has\ rated\ item$} {
    $item\_prediction\ +\hspace{-0.1cm}= abs(s(item, v_j)) * (r_{u_i, item} -
    \overline{r}_{item})$;
    $item\_similartity\ +\hspace{-0.1cm}= abs(s(item, v_j))$;
  }
  $item\_prediction\ /\hspace{-0.1cm}= item\_similarity$;
 }
 $prediction = \overline{r} + user\_prediction + item\_prediction;$
\end{algorithm}

On the other hand, relevance ranking task concerns about how
relevant that the current user and an item are. This problem equals
to ranking the items according to their similarities to the current
user. The detailed algorithm is shown in Alg.
\ref{algo:aca_ranking}.

\begin{algorithm}[!h]
\SetAlgoLined
\KwIn{$u_i, Ph(U), Ph(V)$} \KwOut{Recommendation list}
 \caption{Ranking-based recommendation}
 \label{algo:aca_ranking}
 // Calculate user-item similarities \\
 \For {$item \in I$} {
   $s(u_i, item) = \frac{\mathbf{Ph_{u_i}} \circ \mathbf{Ph_{item}}}{|\mathbf{Ph_{u_i}}| \times |\mathbf{Ph_{item}}|}$;
 }
 // Ranking \\
 $\mathbf{return}$ $Top\ N$ list of items ranked by $s(u_i, item)$ in
 descendant order
\end{algorithm}

\section{Ant Collaborative Filtering with Dimension Reduction}
\label{section4} In the previous section, we have explained the
basic Ant Collaborative Filtering algorithm. Taking one step further
to solve the sparsity problem, we try to improve the ACF by
introducing dimension reduction technique into our previous scheme.

\subsection{Dimension Reduction Version for ACF}
In Algorithm ACF, for every single user we allocate a unique type of
pheromone. Since we intend to represent preference patterns using
different pheromones, much less types of pheromones are actually
needed. We can reduce the number of types of pheromones and thus can
help alleviate sparsity problem. This improved version of ACF is
named IACF.

We first cluster the users into $\mathscr{K}$ clusters according to
their rating pattern, where $\mathscr{K}$ is the desired number of
types of pheromones. Because user clustering is for pheromone
initialization and principally we can use various clustering
techniques \cite{ClusteringSurvey}. We have tested several
clustering methods such as K-means and spectral clustering. We
implemented these two methods and found K-means is more suitable for
our dataset because it generated more balanced clusters. Without
special explantation, we use K-means as initialization process in
our following IACF algorithms.

The detailed IACF algorithm is shown in Alg. \ref{algo:iacf}. The
difference between ACF and IACF is in their pheromone initialization
phase, the rest of the algorithm is kept unchanged as Alg.
\ref{algo:aca_training}.

\begin{algorithm}[!h]
\SetAlgoLined
\KwIn{$Ratings$, $U$, $V$, $\mathscr{K}$} \KwOut{Updated
user pheromone and item pheromone}
 \caption{IACF training algorithm}
 \label{algo:iacf}
 // Clustering users \\
 Cluster users into $\mathscr{K}$ clusters according to their rating
 patterns using K-means; \\
 //Initialization \\
  \For{$user \in U$} {
    // The user belongs to cluster $k$ \\
    $\mathbf{Ph_{user}} = \{Pheromone_{k}: 1.0\}$;
  }
  \For {$item \in I$} {
    $\mathbf{Ph_{item}} = \{\}$;
  }
  // The rest of the algorithm is the same as Alg.
  \ref{algo:aca_training}
\end{algorithm}

\subsection{Complexity Analysis}
For batched training phase the time complexities for both ACF and
IACF are $O(K \times \#ratings)$, where $K$ is the maximum number of
types of pheromones that a user and item carry for both ACF and
IACF. Typically, we have $k<<\#users$. For online update, the update
complexity is only $O(K)$. Generally speaking, IACF is a little
faster because typically there are less types of pheromones attached
to the user and item.

In the recommendation phase, for the rating-based recommendation,
the time complexity is $O(\#users + \#items)$, but the computations
could be significantly reduced if we maintain user neighbors and
item neighbors explicitly in memory or in database. For the
ranking-based recommendation, the time complexity is $O(\#items)$.
This is the lower bound for the recommendation algorithms because we
must scan all the item list before generating a $TopN$
recommendation list.

\section{Related Works}
\label{section5} \vspace{-2.8mm}
\subsection{Bipartite Graph based Collaborative Filtering}
\vspace{-2.8mm} Modeling data mining task as a bipartite graph
analysis problem has a long tradition. \cite{Horting} might be the
earliest work that applied graphic analysis technique in CF
recommendation. Comparing with one mode graph, it is more natural
and precise to model CF problem as a bipartite graph with users and
items as two disjoint groups of nodes. The methods following this
direction CF based on generally fall into three categories: spectral
analysis of the associated matrix, random walks and Activation
Spreading.

\cite{Spectral} introduced bipartite graph co-clustering algorithm
based on the usage the second smallest eigenvector of the associated
matrix of the graph. It achieves the optimal clustering results in
the sense of minimization the cuts of edges between any two
clusters. \cite{WhoRatedWhat} followed this direction and applied
this technique on the task of finding missing links in a movie
rating bipartite graph.

The most well known bipartite graph algorithm in Information
Retrieval may be HITS \cite{HITS} proposed in 1998. It calculates
the stationary status of both groups of nodes through mutual
reinforcement on the bipartite graph. It is a typical random walk
method on the bipartite graph. Similar works include
\cite{RandomWalks} and \cite{Studying}. \cite{Studying} try to find
most relevant users using the user-item Bipatite graph which
recommendations are generated upon. Interestingly,
\cite{SpectralRandom} links the random walks methods and spectral
based methods. Actually, spectral methods are often implemented
using random walks iterations.

Another important technique in graph mining is Activation Spreading
(AS) which models the relationships among the nodes in the graph
through iteratively propagation of activation value.
\cite{ActiveSpreading} surveyed several AS methods and compared them
on a CF task. Their conclusion is that Hopfield net algorithm
outperforms others on their book recommendation data set.
\cite{SpreadingActivation} applied AS technique on rating-based CF
and proposed a novel HITS-like CF algorithm: RSM. We will compare it
with our proposed ACF algorithm in the experiments. In addition, ACF
can also be viewed as a AS extension in recommender systems, the
major difference is that we view the Bipartite user-item
relationship as a dynamic network and learn the training data
incrementally.

The underline principal of the latter two categories is to use
smooth techniques to alleviate the insufficiency of known value of
either the vertices or the edges. This direction is specially
important for collaborative filtering because of sparsity of rating
data.

\subsection{Dimension Reduction in Collaborative Filtering}
Finding a lower rank representation of the original rating data is an
often used way to combat sparsity. The bulk of model based
collaborative filtering are dimension reduction methods in their
essence. The underlying reason is there are much fewer factors than
the dimension of the rating data that actually govern users'
choices. So in practice, dimension reduction can improve both the
performance and efficiency of recommendation algorithms.

Model based recommendation algorithms include probabilistic methods
and matrix factorization methods. The probabilistic methods assume
there are some latent topics that both the users and the items
belong to. The crux is to learn these probabilities that similar
users and similar items are in similar topics. Probabilistic Latent
Semantic Analysis (PLSA) \cite{PLSA} is one of such algorithms.
Another model-based algorithm, Non-negative Matrix Factorization
(NMF) \cite{NMF} belongs to the matrix factorization methods that
also have many extensions and applications in CF domain. By
restricting the rank of the factorized matrices as the user profile
and item profile, these methods are themselves the well known
dimension reduction techniques and are considered to be the
state-of-art of CF algorithms \cite{Temporal}.

\subsection{Swarm Intelligence Recommendation}
To our best knowledge, there are few works that directly related to
our model in the CF domain. But in a more general sense, several
swarm intelligence algorithms have been applied to recommendation
tasks.

\cite{HeatDiffusion} applied heat diffusion model on online
advertisement task. The basic idea is that heat always flows from a
position with high temperature to a position with low temperature.
Influence transmission between friends in a social network is
similar to this diffusion process. \cite{Particle} used particle
swarm optimization algorithm to model every user as a particle in a
multi-dimensional space with each dimension representing a movie
genre. Its advantage is that every particle have a unique position
and velocity, so the system is dynamic and to a certain extent
probabilistic in nature, which is preferable in recommender systems.

\subsection{Comparing with ACF}
Through the analysis above, we can see the relationships between ACF
and other existing methods including bipartite graph based and swarm
intelligence based CF methods especially Activation Spreading
algorithms. The improvements of ACF are three-fold: (1) Comparing
with Activation Spreading algorithms, because of the mechanism of
pheromone evaporation we don't have over-spreading problem which
results in too strong relationship between users and items to find
high quality neighbors; (2) Our training process is performed online
and in accordance with the real time sequence of the user
preferences; (3) Both rating-based and ranking-based algorithms are
considered. For the systems with both two types of preference data
like Amazon\footnote[1]{http://www.amazon.com} and
Douban\footnote{One of the biggest book recommendation websites, we
use its data in our experiment in the following section}, these two
methods can be complement to each other.

\section{Experiments}
\label{section6} As mentioned in Section 2.1, collaborative
filtering methods follow either of two different strategies: rating
prediction and top-N ranking. We experiment our proposed methods on
both of two different scenarios.
\subsection{Rating-based Recommendation}
We experiment the rating prediction algorithms on a popular movie
recommendation dataset: MovieLens
(http://www.grouplens.org/node/73). The MovieLens data we use
consist of $1,000,000$ ratings from $6,040$ users on $3,706$ movies.
Ratings are made on a five-star scale. Of these ratings, $90\%$ are
used for the model training, and the rest $10\%$ constitute test
set.

The evaluation metric is Root Mean Square Error (RMSE) as follows
(the smaller, the better):

$$RMSE=\sqrt{\frac{1}{n}\sum_{(u_i,v_j)\in TestSet}(r_{i,j}-estimated\_rating)^2}.$$

We have some parameters in our algorithm, namely transmission rate
$\gamma$, evaporation rate $\lambda$ and the cluster number for IACF
to be tuned. The transmission rate $\gamma$ controls the speed of
pheromones that are transferred from the user to the item and vice
versa. The bigger, the faster the pheromones transmit. The
evaporation rate $\lambda$ controls the speed of the pheromone
evaporation both on users and items. Bigger value results in slower
evaporation, which means weaker time influences in recommendation.
In our experiments, $\gamma$ is set to be $0.2$ and $\lambda$ is 1.
For IACF, the cluster number is $20$.

We compare our proposed methods with two memory based methods:
classic user-based CF and item-based CF \cite{ItemBased}, two model
based methods: Probabilistic Latent Semantic Analysis (PLSA)
\cite{PLSA}, Non-negative Matrix Factorization (NMF) \cite{NMF} and
an Activation Spreading method: RSM \cite{SpreadingActivation}. The
comparison results are shown in Table \ref{tab:rating}.

\begin{table}[h]
\caption{Rating Results Evaluation} \label{tab:rating}
\begin{center}
\begin{tabular}{p{2.5cm}p{2.5cm}p{2.5cm}} \hline
 \textbf{Algorithm} & \textbf{RMSE} & \textbf{Time (s)} \\ \hline
 User-based & $0.993 \pm 0.02$ &  $98.5$ \\
 Item-based & $0.979 \pm 0.01$ & $1990.0$\\
 NMF & $1.127 \pm 0.05$ & $1605.0$ \\
 PLSA & $1.093 \pm 0.02$ & $102541.0$\\
 RSM & $1.016 \pm 0.02$  &  $581.0$ \\
 ACF & $1.008 \pm 0.03$ & $85.4$\\
 IACF & $\mathbf{0.953 \pm 0.03}$ & $\mathbf{60.2}$\\
 \hline
\end{tabular}
\end{center}
\end{table}
\renewcommand\arraystretch{.6}

Although MovieLens data have timestamp information attached to each
rating, we don't consider time effects because the timestamp don't
reflect the evolving of user interests at all. So all the algorithms
are trained in a batched fashion without special time consequence.
We will see the influences of time in the following experiments.

\subsection{Ranking-based Recommendation}
In order to evaluate the performance of our method on ranking-based
recommendation scenario, we experiment our ranking-based
recommendation algorithm on two real-world recommender systems: book
recommendation and music recommendation. The book recommendation
data are crawled from the largest Chinese book recommendation
website: Douban (http://www.douban.com). For experiment, we use part
of the readers and books as our training/testing dataset. The music
recommendation data are crawled from the largest online music
recommender system: Last.FM (http://www.last.fm). We use these two
datasets because: (1) they are of higher quality than experimental
dataset in terms of reflecting real user preferences; (2) they
symbolize two types of popular recommender systems; (3) they both
contain time information in their implicit ratings that we are
interested in. Table \ref{tab:dataset} shows some statistics of the
datasets.

\begin{table}[h]
\caption{Dataset Properties} \label{tab:dataset}
\begin{center}
\begin{tabular}{p{1.5cm}p{3.0cm}p{3.0cm}} \hline
  & \textbf{Douban} & \textbf{Last.FM} \\ \hline
 \#User & $124$ & $675$ \\
 \#Item & $14843$ & $8010$ \\
 \#Preference & $24862$ & $14007$ \\
 Time Span & \small{$2005.3.7\sim2009.6.22$}& \small{$2006.3.3\sim2009.6.22$}\\
 Sparsity & $98.65$\% & $99.74$\% \\
 \hline
\end{tabular}
\end{center}
\end{table}

However, one of the flaws for ranking-based recommendation is lack
of evaluation metrics as effective as rating-based recommendation.
In order to be less subjective, we use two metrics: Precision
\cite{Ranking} and Ranking Accumulation (RA) \cite{Bipartite}
defined as follows.
$$Precision=\frac{\#\ Hitting}{N},$$
$$RA=\sum_{item \in Hitting} \frac{rank(item)}{N} + \sum_{item \not \in Hitting} \frac{N+1}{N}.$$

Apparently, $Precision \in [0,1]$, the higher the better; $RA \in
[\frac{N+1}{2}, N+1]$, and smaller value is better. (Note that the
$Hitting$ set only contains the items that both are interesting to
the user and already in our test set. So the measured precision and
ranking accumulation both underestimates the real performance.) We
compare our methods with the three ranking-based CF methods reported
in \cite{Bipartite,SpreadingActivation,Ranking} respectively
together with classic user-based and item-based CF algorithms. The
methods proposed in \cite{Bipartite} and \cite{SpreadingActivation}
belong to the general Activation Spreading and are denoted as NBI
and RSM \footnote[1]{The RSM algorithm in the previous section is
the original algorithm, and RSM algorithm in this section is the 0/1
preference implementation}, respectively. We also implemented the
BM25-Item algorithm in \cite{Ranking}. Similarly, we hold $10\%$
data for testing. All the results are obtained by averaging 5
different runs. The comparison results are shown in Table
\ref{tab:ranking} below.

\begin{table}[h]
\scriptsize \small \caption{Ranking Results Comparison}
\label{tab:ranking}
\begin{center}
\begin{tabular}{p{1.5cm}|p{2.5cm}p{1.5cm}p{1.5cm}p{1.5cm}} \hline
\textbf{Dataset} & \textbf{Algorithm} & \textbf{Precision} &
\textbf{RA} & \textbf{Time (s)} \\ \hline
  & User-based & $0.045$ & $20.415$ & $\mathbf{17.5}$\\
  & Item-based  & $0.006$ &  $20.923$  &  $155.7$ \\
  & NBI  & $0.002$ & $20.966$ & $928.0$ \\
  Douban & RSM & $0.035$ & $20.571$ & $27.2$\\
  & BM25-Item  &  $0.014$ & $20.850$ & $202.0$ \\
  & ACF & $0.062$ & $20.320$ & $31.1$\\
  & IACF & $\mathbf{0.065}$ & $\mathbf{20.231}$ & $29.4$\\
 \hline
  & User based & $0.050$ & $20.386$ & $\mathbf{6.4}$\\
  & Item based & $0.040$ & $20.433$  & $84.6$\\
  & NBI  & $0.003$ & $20.957$ & $255.7$\\
  Last.FM & RSM & $0.049$ & $20.356$ & $18.4$ \\
  & BM25-Item  & $0.028$ & $20.726$ & $219.8$\\
  & ACF & $0.076$ & $20.102$ & $25.1$\\
  & IACF & $\mathbf{0.081}$ &  $\mathbf{19.915}$ & $22.3$ \\ \hline
\end{tabular}
\end{center}
\end{table}

\begin{figure}[h]
\centering
\includegraphics[width=90mm,height=50mm]{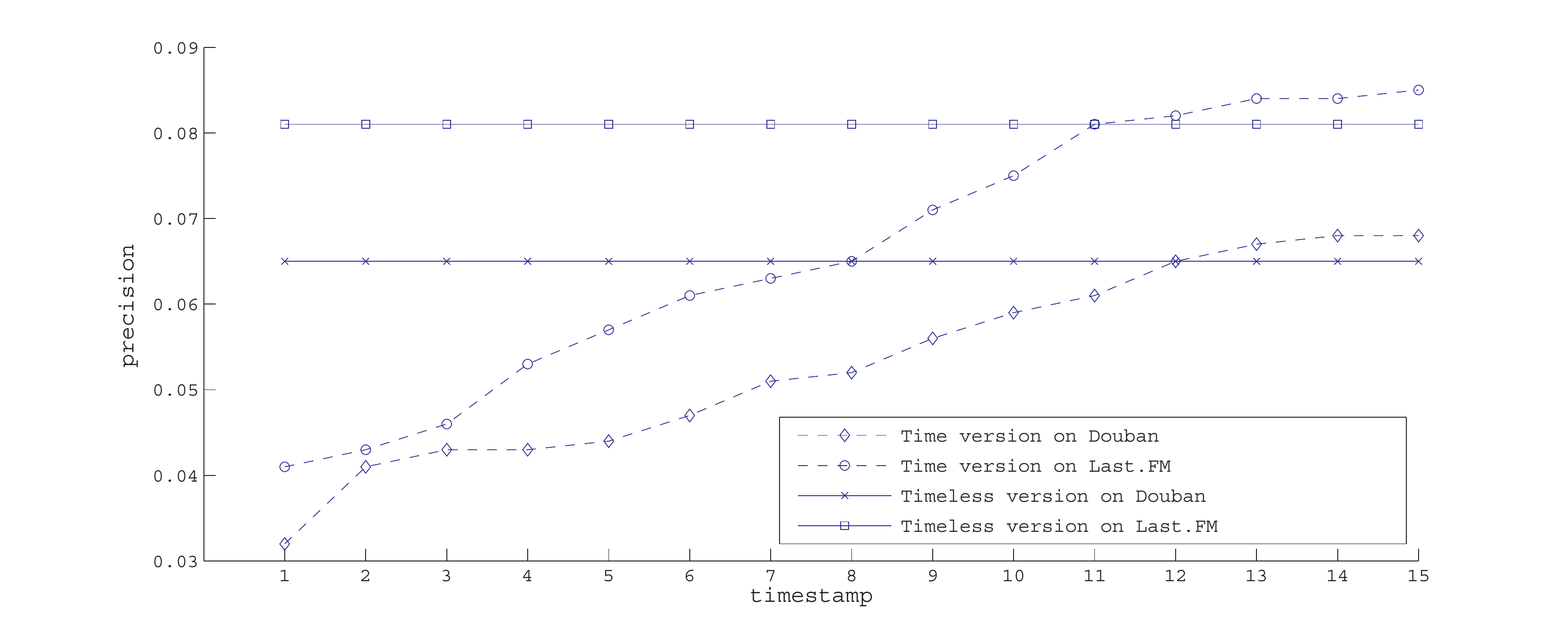}
\caption{Precision comparison of time and timeless versions for IACF
on the Douban and Last.FM dataset} \label{fig:time}
\end{figure}

For all the algorithms in the above experiments we use batched
training data without considering time influence. Since the
personalize recommendations such as book and music recommendation
performances are heavily depending on the time, results can be
further improved by updating our IACF model using preference data in
their original time order. The IACF with and without considering
time sequence are referred to as time and timeless version. We
choose 15 time points within the time span, denoted as $1\sim15$
from $2006.3.3$ to $2009.6.22$. The results are shown in the Fig.
\ref{fig:time}. From the results we could see a clear increase of
precision as time goes on. It means when users keep using our
recommender systems, the recommendation will be more and more
accurate.

\section{Conclusion and Future Works}
\label{section7} Recommendation is becoming of more importance for
many Internet services. We have seen plenty of researches concerning
on improving the accuracy of recommendation algorithms on static,
rating-based data. But as pointed out in the very recent paper
\cite{Temporal}, improved accuracy is not a panacea, there are also
other challenges for collaborative filtering. Scalability is one of
most mentioned concerns and time effects are also a indispensable
factor to be considered in dynamic recommender systems. In this
paper, we proposed a novel CF algorithm inspired by the ant colony
behavior. By pheromone transmission between users and items and
evaporation of those pheromones on both users and items over time,
ACF could flexibly reflect the latest user preference and thus make
most reliable recommendations.

In a nutshell, our major contributions are:
\begin{itemize}
\vspace{-2.8mm}
\item It introduced the concepts in Ant Colony Optimizations (such as Pheromone, Evaporation and
etc.) into recommendation domain and proposed an incremental,
scalable collaborative filtering algorithm that can nicely handle
sparse and evolving rating data;
\item It fused dimension reduction with the Ant Collaborative
Filtering algorithm above mentioned;
\item The algorithm proposed in this paper could recommend with both strategies: rating-based
and ranking-based recommendation which are used in explicit user
preference and implicit user preference scenarios respectively;
\item Last but not least, ACF algorithm is easy to be deployed
on distributed computational resources, even in a Peer-to-peer
environment, which means a higher scalability and more importantly,
user privacy protection.
\end{itemize}

There are also some unexplored possibilities to improve the
algorithm proposed in this paper. First, the initialization of
pheromones does affect the final recommendation results as shown in
the dimension reduction version of ACF. There may be other
initialization schemes that we can make further improvements.
Second, evaporation is an interesting while hard-to-tune mechanism
that applications should find the most suitable rate to their own
needs. Last but not least, Bipartite based ranking methods including
ours are flexible and can often fuse user preference data other than
user ratings \cite{Co-HITS}. This is also a promising direction.

\section{Acknowledgements}
We thank State Key Laboratory of Computer Science for subsidizing
our research under the Grant Number: CSZZ0808, the National Science and Technology Major Project under Grand Number:2010ZX01036-001-002-2 and the Grand Project Network Algorithms and Digital Information of the Institute of Software, Chinese Academy of Sciences under Grant Number:YOCX285056.

\nocite{ex1,ex2}
\bibliographystyle{latex8}

\end{document}